\documentclass[12pt]{iopart}

\usepackage{color}
\usepackage{graphicx}

\newcommand{\nc}{\newcommand}
\nc{\eq}{\begin{equation}}
\nc{\eeq}{\end{equation}}
\nc{\eqa}{\begin{eqnarray}}
\nc{\eeqa}{\end{eqnarray}}
\nc{\ar}{\begin{array}}
\nc{\ear}{\end{array}}
\nc{\bfig}{\begin{figure}}
\nc{\efig}{\end{figure}}
\nc{\dg}{\dagger}
\nc{\sx}{\sigma_x}
\nc{\sy}{\sigma_y}
\nc{\sz}{\sigma_z}
\nc{\spl}{\sigma_+}
\nc{\sm}{\sigma_-}
\nc{\nn}{\nonumber}
\nc{\noi}{\noindent}
\nc{\adg}{a^{\dg}}
\nc{\kvec}{\mathbf{k}}

\begin{document}

\title{Robust non-Markovianity in ultracold gases}

\author{P~Haikka$^{1}$, S~McEndoo$^{1,2}$, G~De Chiara$^{3}$, G M ~Palma$^{4}$ and S~Maniscalco$^{1,2}$}

\address{$^1$ Turku Center for Quantum Physics, Department of
Physics and Astronomy, University of Turku, FIN20014, Turku,
Finland}

 \address{$^2$ SUPA,
EPS/Physics, Heriot-Watt University, Edinburgh, EH14 4AS, United Kingdom}

\address{$^3$ Centre for Theoretical Atomic, Molecular and Optical Physics, School of Mathematics and Physics, Queens University, Belfast BT7 1NN, United Kingdom}

\address{$^4$ NEST Istituto Nanoscienze-CNR and Dipartimento di Fisica, Universit\`a degli Studi di Palermo, via Archirafi 36, I-90123 Palermo, Italy}

\begin{abstract}
We study the effect of thermal fluctuations on a probe qubit interacting with a Bose-Einstein condensed (BEC) reservoir. The zero-temperature case was studied in [Haikka P \etal 2011 \PR A {\bf84} 031602], where we proposed a method to probe the effects of dimensionality and scattering length of a BEC based on its behavior as an environment. Here we show that the sensitivity of the probe qubit is remarkably robust against thermal noise. We give an intuitive explanation for the thermal resilience, showing that it is due to the unique choice of the probe qubit architecture of our model.
\end{abstract}

\maketitle

\section{Introduction}

Over the past three decades quantum computing has been the holy grail of quantum information sciences \cite{nielsen}. Lately, there has been a notable shift of focus from studies of circuit based quantum computers, where a long computation involving many qubits is broken down to elementary one- and two-qubit quantum gates, to studies of quantum simulators, where a physical system is modelled using another physical realisation of the original Hamiltonian \cite{simulations}. One celebrated example of the latter is the simulation of a Bose-Hubbard model in optically trapped ultracold gases. The mapping between the Bose-Hubbard Hamiltonian and the Hamiltonian describing ultracold atoms in an optical lattice was proposed in 1998 by D. Jaksch \etal \cite{dieter} and realized experimentally a few years later by the group of I. Bloch \cite{bloch}. Since this milestone there has been an explosion in studies of systems that can be simulated with ultracold quantum gases \cite{sanpera review, bloch review}.

Quantum simulations have been considered in the context of open quantum systems with proposals of simulating the spin-boson model using, for example, a quantum dot coupled to a Luttinger liquid \cite{recati} or a more general Bose-Einstein condensed (BEC) reservoir \cite{dieter2}. Both cases realise the independent-boson Hamiltonian with an Ohmic-like spectrum of the reservoir \cite{independent boson}. Another proposal in this direction was presented in Ref. \cite{massimo}, where an impurity atom in a double well potential is immersed in a BEC reservoir, forming a spin-boson model with a reservoir spectral function that can be tuned from sub-Ohmic to Ohmic to super-Ohmic. With a super-Ohmic spectrum the spin-boson model can acquire non-Markovian properties \cite{us}, hence simulating a prototype of a non-Markovian open quantum system model.

Non-Markovian systems have been the subject of intense studies in recent few years, boosted by the recent introductions of several non-Markovianity measures that define and quantify the amount of non-Markovianity in a quantum process \cite{BLP, RHP, fisher}. Fundamental interest in non-Markovian processes stems from the fact that Markovian dynamics is typically only an approximation, which is no longer valid when considering shorter time-scales and/or stronger system-environment couplings. Furthermore, there have been proposals for using non-Markovianity as a resource in the context of quantum metrology \cite{huelga} and quantum key distribution \cite{ruggero}, to name a couple of examples.

Spin systems coupled to ultracold gases are not only important for quantum simulations, but also because they can be used to probe the ultracold gas: the way a spin decoheres under the action of the ultracold gas may depend crucially on certain properties of the gas. Hence it is possible to recover information about the large and generally inaccessible environment by looking at the spin only. Indeed, the aforementioned independent boson models can be used to probe the Luttinger liquid parameter \cite{recati} and the density fluctuations of the BEC \cite{dieter2}. In Ref. \cite{us} we demonstrated that the non-Markovian properties of the impurity atom in a double well potential give indications of the effective dimensionality of the BEC reservoir. In this work we further consider this model, taking a step towards a more realistic scenario by considering the effect of thermal fluctuations on the sensitivity of the probe qubit. We demonstrate that the double well qubit is remarkably robust against thermal noise and therefore a good candidate for probing ultracold gases.

\section{The model}

We consider a qubit model based on a single atomic impurity trapped in a double well potential, where the pseudo-spin states are represented by presence of the impurity atom in the left or the right well of the double well potential. The qubit is immersed in a thermally equilibrated Bose-Einstein condensate reservoir. The Hamiltonian of the total closed system is ($\hbar=1$)
\eq \label{Ham}
	\hat{H} = \hat{H}_A + \hat{H}_B + \hat{H}_{AB},
\eeq
where
\eq \label{HA}
	\hat{H}_A = \int d{\bf x}\; \hat{\Psi}^\dagger({\bf x}) \left[ \frac{{\bf p}^2_A}{2 m_A} + V_A({\bf x}) \right] \hat{\Psi}({\bf x})
\eeq
is the Hamiltonian of the impurity atom with $\hat{\Psi}({\bf x})$ the impurity field operator and $V_A({\bf x})$ the double well potential formed by an optical lattice,
\eq  \label{HB}
	\hat{H}_B = \int d{\bf x}\; \hat{\Phi}^\dagger({\bf x}) \left[  \frac{{\bf p}^2_B}{2 m_B} + V_B({\bf x}) + \frac{g_B}{2}  \hat{\Phi}^\dagger({\bf x})\hat{\Phi}({\bf x})  \right] \hat{\Phi}({\bf x})
\eeq
is the Hamiltonian for the BEC with $\hat{\Phi}({\bf x})$ the condensate field operator, $V_B({\bf x})$ the harmonic trapping potential and $g_B = 4\pi \hbar^2a_B/m_B$ the boson-boson coupling constant, and finally
\eq  \label{HAB}
	\hat{H}_{AB} = g_{AB} \int d{\bf x}\; \hat{\Psi}^\dagger({\bf x})   \hat{\Phi}^\dagger({\bf x}) \hat{\Phi}({\bf x})  \hat{\Psi} ({\bf x})
\eeq
is the interaction Hamiltonian with $g_{AB} = 2\pi\hbar^2 a_{AB}/m_{AB}$ the coupling between the impurity atom and the condensate gas. Masses of the impurity atom and the background bosons are $m_{A/B}$, $m_{AB}=(m_A+m_B)/(m_Am_B)$ is their reduced mass and $a_{B/AB}$ are the s-wave scattering lengths for the boson-boson and impurity-boson collisions, respectively. 

We would like to stress that the boson-boson scattering length can be manipulated by Feshbach resonances, providing a controllable environment of interacting bosons. The significance of this is twofold: for one, many typical models of open qubit systems assume a non-interacting bosonic environment and it is fundamentally interesting to study interacting models. Secondly, the ability to have experimentally feasible and precise control over the environment is vital for reservoir engineering.

We assume that the BEC is trapped in such shallow potential that it may be considered to be homogenous, while the double well trap for the impurity atom is so deep that tunneling from one well to the other is suppressed. The condensate is treated in the Bogoliubov approximation, assuming weak to moderate boson-boson coupling. After imposing these assumptions on Hamiltonians  \eref{HA}-\eref{HAB} the qubit dynamics can be derived without any further approximation. The result is purely dephasing dynamics of the qubit with constant populations and off diagonal elements of the qubit density matrix decaying as
\eq
	|\rho_{01}(t)| =  e^{-\Gamma(t)} |\rho_{01}(0)|.
\eeq
The decoherence factor is
\eq \label{decoh}
	\Gamma(t) = 8 g_{AB}^2 n_0 \sum_{\mathbf k} (|u_k| - |v_k|)^2 e^{-k^2 \tau^2 /2} \frac{\sin^2( E_k t/2\hbar)}{E_k^2} \coth\left( \frac{\beta E_k}{2}\right) \sin^2 ({\mathbf k} \cdot {\mathbf L}),
\eeq
where $n_0$ is the condensate density, $|u_k|$ and $|v_k|$ are the $k$-th Bogoliubov mode amplitudes with energy $E_k = \sqrt{ 2 \epsilon_k n_0 g_B + \epsilon_k^2}$, free modes have energy $\epsilon_k = \hbar^2 k^2 /(2m_B)$, $\tau$ is the width of the impurity wavefunction, assumed Gaussian, in each well of the double well and $\beta = 1/k_B T$. The spatial separation between the two wells is $\mathbf{L}$. For a detailed derivation of the decoherence factor see \cite{massimo, us}.

\section{Non-Markovianity Measure}
With the decoherence factor at hand one has a full description of the qubit dynamics. We proved in Ref. \cite{us} that in this case non-Markovianity is directly connected to the negativity of the decay rate $\gamma(t) = d \Gamma(t) / dt$. In this Section the connection is briefly reviewed and extended to the case of thermal reservoirs. We focus on the approach of Ref. \cite{BLP}, which defines Markovianity to be a property of a dynamical map $\rho(0)\mapsto\rho(t)=\Phi(t,0)\rho(0)$ that monotonically decreases the distinguishability $D[\rho_1,\rho_2]=\frac{1}{2}|\rho_1-\rho_2|$ of any two system states $\rho_{1,2}(t)$. Non-Markovianity is then the ability of a dynamical map to temporarily increase the distinguishability of two states. The temporal change in the distinguishability $\sigma=dD[\rho_1,\rho_2,t]/dt$ can be associated to information flowing from the system to its environment ($\sigma<0$)  or back to the system ($\sigma>0$). The amount of non-Markovianity in a quantum process is given by the cumulant of the positive information flux, $\mathcal{N}=\max_{\rho_1,\rho_2}\int_{\sigma<0}ds\,\sigma(s)$, with a maximization done over all possible pairs of states to find the largest amount of information that the system can recover from the environment.

The maximization required in the calculation of the non-Markovianity measure is generally difficult. In the case of pure qubit dephasing, however, it has been proven that the optimising pair is formed by two antipodal states in the equator of the Bloch sphere and in this case the measure can be recovered analytically. One finds easily that information flows back to the qubit from the environment iff the decay rate is negative. Moreover, in the model considered in this article there is at most a single interval of time $a<t<b$ such that $\gamma(t)<0$ and thus we introduce a normalized version of the non-Markovianity measure, which measures the amount of recovered information against the amount that was lost to the environment in the time interval $0<t<a$. Summarizing, the measure we use in this work to study the non-Markovian properties of a qubit dephasing in a BEC environment is
\eq
	\mathcal{N} = \frac{e^{-\Gamma(b)} - e^{-\Gamma(a)}} { e^{-\Gamma(0)} - e^{-\Gamma(a)}}, \quad \gamma(t)=\frac{d\Gamma(t)}{dt}<0\Longleftrightarrow t\in[a,b].
\eeq
In Ref. \cite{us} we studied the changes in the non-Markovianity measure induced by different effective dimensions of the reservoir and for a range of different values of the scattering length of the environment, assuming a zero-T environment.  We found the existence of a dimension-dependent critical scattering length such that when $0\leq a_B<a_{crit}$ the dynamics  of the qubit is Markovian and when $a_B>a_{crit}$ it is non-Markovian. The dependence on the effective dimension of the BEC is such that $a_{crit, 3D}<a_{crit, 2D}<a_{crit, 1D}$, that is, the higher is the dimension the smaller is the critical scattering length. Hence one has at hand a model where the Markovian to non-Markovian crossover can be controlled by changing either the scattering length of the background bosons or by lowering the effective dimension of the BEC. Conversely, one may deduce these properties of the environment by looking at the qubit only, without directly measuring the BEC. In this work we proceed to consider the effect of thermal fluctuations on this result. Thermal fluctuations can, in principle, wash out non-Markovian effects and thus compromise the sensitivity of the probe qubit. Fortunately we find the double well qubit model to have a remarkable robustness against thermal effects, as shown in the next section.

\section{Results}
\begin{figure}[t]
\begin{center}
\includegraphics[width=0.8\linewidth]{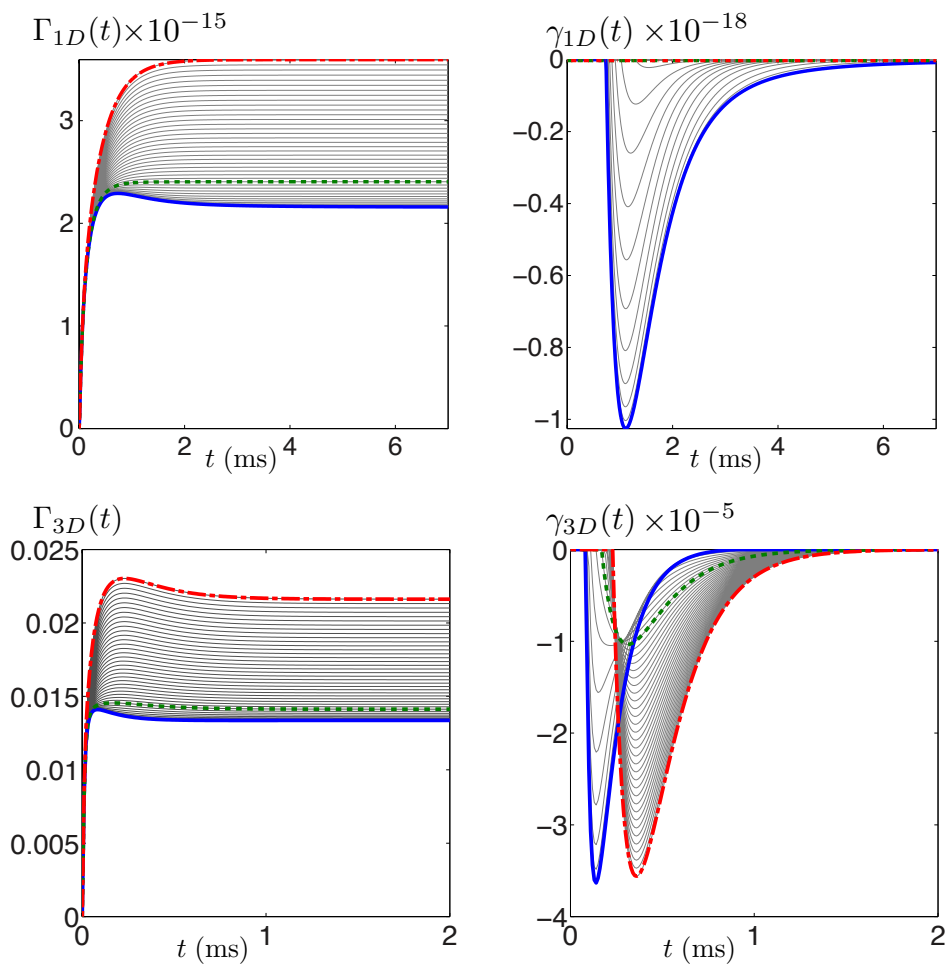}
\end{center}
\caption{(Color online) Top: Decoherence factor, $\Gamma_{1D}(t)$ and decay rate, $\gamma_{1D}(t)$ for a  one-dimensional environment, for temperature ranging between $0$ K (solid blue line) and $20$ nK (dotted red line).  The $T = 6.5$ nK line, shown in dashed green, show the  transition from non-Markovian low temperature limit and the Markovian high temperature limit.\\
Bottom:   Decoherence factor, $\Gamma_{3D}(t)$ and decay rate, $\gamma_{3D}(t)$ for a three-dimensional environment, for temperature ranging between $0$ K (solid blue line) and $200$nK (dotted red line). The $T = 50$ nK line, shown in dashed green, shows the intermediate stage between the low and high temperature limits.}
\label{temp_3D}
\end{figure}

Fig.~\ref{temp_3D} shows the decoherence factor $\Gamma(t)$ and decay rate $\gamma(t)$ for a three-dimensional and a one-dimensional $^{87}$Rb condensate with a range of temperatures $T=0$ to $200$ nK and $T = 0$ to $20$ nK respectively. We take the same parameters as in Ref. ~\cite{us} and a fixed value $a_B=a_{Rb}$ for the scattering length. In the 1D case the negative part of the decay rate, indicating the existence of non-Markovian effects, decreases in size with increasing temperature until it reaches a critical temperature of around  $T = 6.5$ nK, where it vanishes completely. When this happens, the qubit dynamics is Markovian. In the 3D case the negative part of the decay rate splits into two lobes: the low temperature lobe, enclosed by the line corresponding to the zero-T decay rate, and the high temperature lobe, indicated by the high temperature decay rate. The transition between the two lobes, indicated by the decay rate for $T = 50$ nK, corresponds to the transition from the low temperature regime to the high temperature regime. We demonstrate next that the transition from low to high temperatures is also clearly visible in the non-Markovianity measure.

\begin{figure}[t]
\begin{center}
\includegraphics[width=0.7\linewidth]{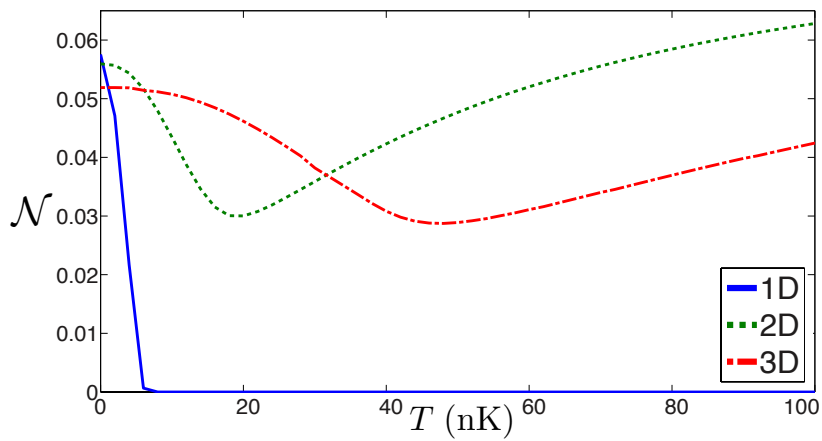}
\end{center}
\caption{(Color online) Non-Markovianity measure $\mathcal{N}$ as a function of temperature for a quasi-1D (blue solid), quasi-2D (green dashed) and 3D (red dotted) environments. Scattering length is fixed at $a_B = a_{Rb}$.}
\label{temp}
\end{figure}

The temperature dependence of the non-Markovianity measure $\mathcal{N}$ is shown in Fig.~\ref{temp} for all three effective dimensions. In the case of a quasi-1D condensate, the system is non-Markovian only for very small temperatures $T\sim1$ nK and for a higher temperature thermal fluctuations wash out memory effects in the qubit dynamics. When the qubit is embedded in a quasi-2D or a 3D condensate, the dynamics is more robust against thermal effects. In these cases the non-Markovianity measure is almost constant for low temperatures when $\coth(\beta E_k/2)\approx1$. As the temperature is raised, the value of the non-Markovianity measure decreases as the system moves towards the high temperature regime, while in the high-T regime the measure increases in value again. The minima in Fig. 2 corresponds to the decay rate moving from the low-T lobe to the hight-T lobe (see Fig. 1). In the high-T regime $\coth(\beta E_k/2)\approx(\beta E_k/2)^{-1}$, and temperature acts as a coefficient for the decoherence factor of Eq. \eref{decoh}. Consequently the whole dynamics is amplified, leading also to higher values of the non-Markovianity measure.

\begin{figure}[t]
\begin{center}
\includegraphics[width=0.8\linewidth]{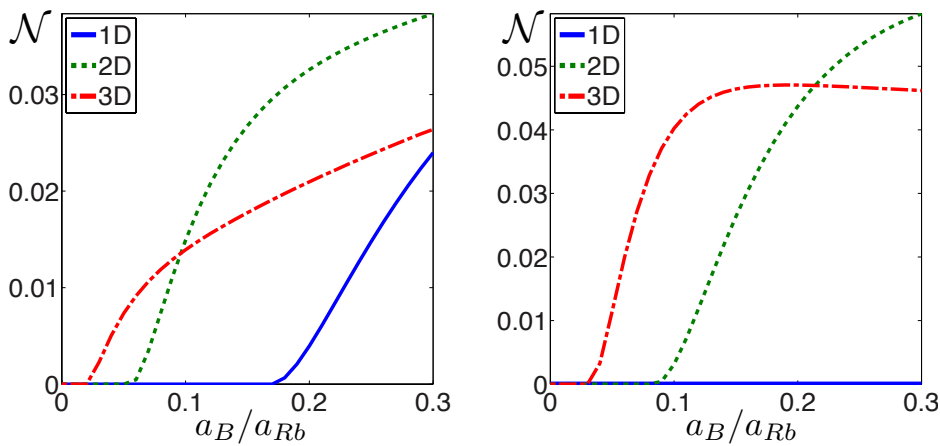}
\end{center}
\caption{(Color online) Non-Markovianity measure $\mathcal{N}$ as a function of relative scattering length $a_B/a_{Rb}$ for  quasi-1D, quasi-2D and 3D environments with temperatures $T = 0.5$ nK and $T = 100$ nK.}
\label{scattering}
\end{figure}

Finally, Fig.~\ref{scattering} shows the non-Markovianity measure $\mathcal{N}$ as a function of the manipulated scattering length of a  Bose-Einstein condensate for temperatures $T=0.5$ nK and $T=100$ nK. In both cases we reproduce the main result of Ref. \cite{us}, namely that the qubit system has a transition from Markovian to non-Markovian dynamics with increasing scattering length, and that the critical scattering length depends on the effective dimensionality of the BEC: $a_{crit, 3D}<a_{crit, 2D}<a_{crit, 1D}$. For high temperatures the quasi-1D environment is unable to return information back to the system, leading to purely Markovian dynamics. However, since the quasi-2D and the 3D environments still induce non-Markovian dynamics, the information obtained on the effective dimensionality of the environment by looking at the qubit dynamics is the same. Thus we find our main result: thermal effects do not compromise the ability of the probe qubit to detect information about the effective dimensionality and the scattering length of the environment.

\section{Discussion and Conclusions}
The remarkable sensitivity of the probe qubit we propose in this paper derives from its general robustness against thermal effects. This, in turn, is due to the very specific qubit architecture we choose. The deep double well potential, in which the impurity atom is trapped, imposes limitations on the contribution of certain Bogoliubov modes to the qubit dynamics. The decoherence function $\Gamma(t)$ is defined as an integral over all modes $k$, however in this model there are two cut-off momenta: $1/\tau$, relates to the size of each harmonic potential in the double well and $1/L$ characterizes to the distance between the minima of the wells, and only excitations corresponding to $1/L < k < 1/\tau$ contribute to the dynamics. For high temperatures the temperature-dependent term $\coth( \beta E_k/2)$ diverges at $E_k = 0$, that is, when $k=0$. However, the lower cut-off frequency excludes the diverging terms and thus prevents non-Markovianity being washed out in the higher temperature regimes. This is a feature specifically due to the spatial nature of the double well qubit, making this model well suited to realistic temperatures. 

In summary, the double well probe qubit model provides an ideal system for observing non-Markovianity in an atomic system, and for exploiting the Markovian to non-Markovian crossover to probe the BEC environment. In addition to involving systems that are straightforwardly combined experimentally, it is an example of a system showing a Markovian to non-Markovian crossover in accessible parameter ranges. We have shown here that in addition to the above advantages, this system is also robust to temperature effects, with the measured quantity maintaining its size for temperatures up to and, in some cases, beyond those necessary for experimental realization of these systems.

\section*{Acknowledgments}
This work was supported by the Emil Aaltonen foundation and the Finnish Cultural foundation.

\end{document}